## RESEARCH WORKING PAPER

# Cooperation among competitors in the open-source arena
## The case of OpenStack[1]

Jose Apolinario Teixeira[1*], Salman Qayyum Mian[2] and Ulla Hytti[1]

[*]Correspondence:
jose.teixeira@utu.fi
[1]Turku School of Economics, University of Turku, Turun yliopisto, 20014 Turku, Finland
Full list of author information is available at the end of the article

**Abstract**

Interorganizational interactions are often complex and paradoxical. In this research, we transcend two management paradoxes: *competition versus cooperation* and *open-source versus proprietary* technology development. We follow the OpenStack open-source ecosystem where competing firms cooperate in the joint-development of a cloud infrastructure for big data. We provide a narrative, complemented with social network visualizations, which depicts the evolution of cooperation and competition. Our findings suggest that development transparency and weak intellectual property rights (i.e., characteristics of open-source ecosystems) allow a focal firm to transfer information and resources more easily between multiple alliances.

## I. Introduction

ORGANIZATIONS do not live in isolation; they are networked in nature. Many firms, particularly in the high-tech industry, have increasingly engaged in different kinds of cooperative interorganizational relationships (e.g., contractual alliances, joint ventures, consortia, standards bodies and open-source communities) to improve their resource endowment and manage strategic uncertainty (Chen and Miller 2015; Eisenhardt 1989; Hoffmann 2007; Morgan et al. 2013). Consequently, organizations are embedded in dynamic and cohesive networks where individual and organizational actions are influenced by both their network and their network position (Gomes-Casseres 1996; Granovetter 1973; Gulati and Gargiulo 1999; Uzzi 1996).

Scholars early recognized interorganizational relationships as an important source of competitive advantage (Dyer and Singh 1998; Hoffmann 2007; Zaheer et al. 2000). While interorganizational interactions have become more complex, they have also become more paradoxical (Lewis 2000). Paradox pervades daily life, and how to manage paradox has long been the subject of philosophical and organizational inquiry (Chen 2008). Organizational researchers have previously highlighted the significance of paradox in business practice, "excellent companies have learned how to manage paradox" (Peters and Waterman 2004, p.100). Poole and Van de Ven (1989, p.563) urged researchers to seek out forms of paradox and look for theoretical tensions or oppositions to stimulate the development of more encompassing theories (Smith and Lewis 2011).

---

[1]As accepted for presentation at the 2016 International Conference on Information Systems (ICIS 2016), held in Dublin, Ireland, December 11-14, 2016. "IS in Organizations and Society" track. The official conference proceedings are available at the AIS eLibrary (`http://aisel.aisnet.org/icis2016/`).

In this research, we transcend two management paradoxes filled with theoretical tension: *competition versus cooperation* and *open-source versus proprietary* technology development. As a particularly vexing organizational paradox, *competition versus cooperation* occupied historically a central position in management research (Chen 2008; Yami et al. 2010). With a stronger emphasis on R&D and innovation, the paradox of *open-source versus proprietary* technology development has attracted wide multi-disciplinary attention (Crowston et al. 2012; Raasch et al. 2013; von Hippel and von Krogh 2003).

On the one hand, much literature addressing *competition versus cooperation* stresses the concepts of intellectual property, cross-licensing, secrecy and gatekeeping (Bengtsson and Kock 2014; Gnyawali and Park 2009; Ritala and Hurmelinna-Laukkanen 2013). On the other hand, *open-source versus proprietary* technology development stresses the concepts of freedom, transparency, openness and inclusiveness (Aksulu and Wade 2010; Bonaccorsi et al. 2006; Gacek and Arief 2004; Raymond 1999; von Hippel and von Krogh 2003).

In this research, we explore "open-coopetition", a neologism recently proposed by Teixeira and Lin (2014) to describe cooperation among competitors in the open-source arena. There are several known cases of open-coopetition as captured in Table 1. Cooperation among competitors in an open-source way has become common in different R&D intensive sectors – it can be observed in the development of web, cloud computing, mobile, automotive and medical technologies among others.

| Project | Domain | Examples of competing firms cooperating in the project |
|---|---|---|
| WebKit | Web browsing technologies | Apple, Nokia, Google, Samsung, Intel and BlackBerry |
| Blink | Web browsing technologies | Google, Opera, Intel and Samsung |
| OpenStack | Cloud computing infrastructure | Rackspace, Canonical, IBM, HP, Vmware and Citrix |
| CloudStack | Cloud computing infrastructure | Citrix, SunGard AS and ShapeBlue |
| Xen | Virtualization technologies | University of Cambridge, Citrix, IBM, HP and Red Hat |
| Hadoop | Distributed computing technologies | Facebook, Twitter, LinkedIn, Jive, Microsoft and Hortonworks |
| Open Handset Alliance | Mobile devices platform | Google, LG, Samsung, HTC, Huawei, ZTE, Lenovo and NEC |
| Tizen | Mobile devices platform | Fujitsu, Huawei, NEC, Casio, Panasonic and Samsung |
| GENIVI Alliance | In-Vehicle Infotainment platform | Volvo, BMW, Honda, Hyundai, Renault and PSA |
| Linux | The Linux operating system | Fujitsu, HP, IBM, Intel, Samsung, Hitachi and Red Hat |
| Yocto project | Development tools for embedded Linux | Broadcom, AMD, Texas Instruments and Intel |
| Linaro | Development tools for embedded Linux | ARM, Samsung, ST-Ericsson and Texas Instruments |
| Eclipse | Software development environment | Actuate, CA, IBM, Google, Oracle, SAP and Red Hat |
| OpenEMR | Health records and medical practice software | OEMR, EnSoftek, MI-Squared, ZH Healthcare and Visolve |

Table 1 Known cases of cooperation among competitors in the open-source arena (i.e. open-coopetition)

Even if cooperation among competitors and open-source software are phenomena with recognized impact on how value is created, explored and exploited in networked settings, there are very few studies addressing how rival firms simultaneously cooperate and compete in the open-source arena (Germonprez et al. 2013;

Teixeira et al. 2015). From a practitioner's viewpoint, this is unfortunate since naive assumptions concerning "work with competitors" and "open-source work" can lead in practice to opportunistic behavior, unintended spillover effects, and loss of reputation and trust among partners (see Markus and Agres 2000; Nooteboom et al. 1997; Park and Russo 1996; Sharma et al. 2002). Given the scarcity of theory and empirical studies addressing this new phenomenon, we conducted this exploratory case study guided by the broad and open research question: "'how competitors cooperate in open-source ecosystems?".

## II.   Theoretical background

### Strategic networks and ecosystems

Even if many see firms as distinct and autonomous units of action, it has been both sighted and increasingly recognized that firms are embedded in networks comprised of close, robust and multidimensional connections that that blur organizational boundaries (Gilsing et al. 2008; Granovetter 1973; Powell and Smith-Doerr 1994; Rowley et al. 2004; Uzzi 1996). Organizational theory now stresses the need to understand how the relational context in which firms are embedded influences their behaviors (Gulati 1998; Rowley et al. 2004).

Early identified drivers of inter-firm cooperation include: reducing costs through product rationalization and economies of scale (Contractor and Lorange 1992; Mariti and Smiley 1983), sharing total risks and total costs of large projects (Baldi 2012; Harrigan 1988), accessing new knowledge and complementary assets (Kogut 1988; Teece 1992), developing technology and accessing complementary markets (Cravens et al. 1996), co-creation/co-production of value (Chan et al. 1997; Ramirez 1999), solving market failures that emerge under conditions of bounded rationality (Williamson 1985,9), shaping competition with the aim of increasing or decreasing market competition (Fuller and Porter 1986; Garud and Kumaraswamy 1993), faster access to new markets (Contractor and Lorange 1988; Hung et al. 2003), gaining legitimacy and reputation (DiMaggio and Powell 2000; Greenwood and Meyer 2008), securing investment (Guiso et al. 2004; Stam and Elfring 2008) and reducing uncertainty from resource requirements (Pfeffer and Salancik 2003). Even if most interorganizational research takes the point of view of a focal firm embedded within one alliance it is known that uncertainty drives firms to establish a portfolio of alliances – firms are often embedded in multiple strategic networks (Dreyfus et al. 2005; Hoffmann 2007; Lavie and Miller 2008; Lavie and Singh 2012).

The ecosystem construct, as a way of making the networked interdependencies of the firm more explicit, has gained prominence in both research and practice (Adner 2006; Iansiti and Levien 2004a; Moore 1999). Theory underlying the ecosystem construct have focused on understanding coordination among partners in exchange networks that are characterized by simultaneous cooperation and competition (Afuah 2000; Brandenburger and Nalebuff 2011). Studies in this arena have explored the challenges that arise when incentives across the ecosystem are not aligned (Casadesus-Masanell and Yoffie 2007), the role of established relationships with ecosystem partners in shaping firms motivations to compete for different market segments (Christensen and Rosenbloom 1995), the activities that focal firms undertake to induce exchange partners to favor their specific technology platforms (Gawer and Cusumano 2002), and the flow of activity among partner firms (Adner and Kapoor 2010).

In our view, the ecosystems construct stresses the importance of actor-to-actor networked relationships. Therefore, our view on strategic networks (Gulati 1998; Jarillo 1988; Rusko 2014; Zaheer et al. 2000) and ecosystems (Adner 2006; Iansiti and Levien 2004a; Moore 1999) approximates what others called "networked collaborations" (Normann and Ramirez 1993); "lattices" (Gore 1985), "webs" (Hastings 1993), "constellations" (Normann and Ramirez 1994), "holonic organizations" (McHugh et al. 1995), "interfaces" (Gilmore and Krantz 1991), "organizations networks" (Perrow 1972), "inter-organizational domain" (Trist 1977) and "infrastructure" (Tilson et al. 2010).

Prior research has considered mainly the independent motivations and opportunities that guide alliance formation at the dyad-level (Lavie and Singh 2012). In this study, we explore the evolution of cooperation among competitors at multiple levels; we examine relational interactions a the inter-individual, inter-firm and inter-ecosystem level (i.e., zooming in and out). As pointed out by Ibarra et al. (2005), distinctive issues concerned with the alignment of individual and collective networking interests should not be separated.

## Cooperation among competitors

As pointed out by the literature on business ecosystems, strategic cooperation among competitors (aka coopetition) in not uncommon (Clarysse et al. 2014; Iansiti and Levien 2004b). The phenomenon can be found, for instance in the automotive industry. The city car models Toyota Aygo, Peugeot 108 and Citroën C1 share the same body and equipment. In fact, they are made in the same factory, as a result of a joint venture between Toyota Motor Corporation and PSA Peugeot Citroën. In the pharmaceutical industry, 10 giants leading the industry founded TransCelerate BioPharma "as a nonprofit, precompetitive drug company, to develop shared industry clinical-trial solutions". Also in the airline industry, dyadic alliances and multilateral alliances have blurred the borders between airline companies over the last two decades (Gudmundsson and Lechner 2006). Competition between airlines is less a matter of individual firms competing against individual firms but rather of airline alliances against airline alliances (Gomes-Casseres 1994). In the computer and mobile-devices industry, it is also known that Apple, Google and Samsung among others, have cooperated in the development of open-source web-browsing technologies, even while fighting patent-wars (Teixeira and Lin 2014).

On one hand, economists have addressed the phenomenon through public and macroeconomic perspectives. For example, Ausubel (1991); Rochet and Tirole (2002); Schmalensee (2002) have addressed cooperation among competitors in ATM and card-payment payment systems. Much of this work, follows the neo-classical theory in economics in assuming that competition generates economic efficiency (Lado et al. 1997). Antitrust and regulatory policies prohibit many agreements or practices that restrict free trading and competition between businesses (European Commission 2001; Federal Trade Commission 2000).

In parallel, research on alliances within strategic management gives "important insight into the advantages that can be obtained by cooperation and the prerequisites needed for an alliance to work, but it is primarily the cooperative dimension of the relationship that is emphasized" (Bengtsson and Kock 2000). More recently, Gulati et al. (2012) claimed that important streams of research on alliances remain single-mindedly focused on the cooperation perspective. Summing up, from the individual and firm perspective, competition is not necessarily socially desirable (Loury 1979). However, from a public and macroeconomics perspectives, the existence of competition is essential for welfare (Pigou 2013).

Although competition and cooperation have individually received much consideration, given limited attention to the fundamental issue of the interplay between the two concepts (Chen 2008; Chen and Miller 2015). Even if strategic management literature noted the importance of understanding open-source software from its competitive-cooperative angles (see Bengtsson et al. 2010; Chen and Miller 2015; McGaughey 2002), very few empirical cases exploring cooperation among competitors in open-source arena exist. A few notable exceptions here are the studies by Germonprez et al. (2013); Linåker et al. (2016); Teixeira and Lin (2014); Teixeira et al. (2015).

## Open-source software under a network perspective

Much of the innovative programming that powers software applications, operating systems, clouds servers and the Internet is the result of "open-source" code – that is, code that is freely distributed as opposed to being kept secret.

Consistently across recent reviews (Aksulu and Wade 2010; Crowston et al. 2012; Teixeira and Baiyere 2014), there is a general consensus that "open-source" (also known as free-software or software libre) emerged with a set of four freedoms as suggested by Stallman (1985). These freedoms laid down the foundations for the open-source software as known today: 1) the freedom to run the program, for any purpose; 2) the freedom to study how the program works, and change it so it does your computing as you wish; 3) the freedom to redistribute copies so you can help your neighbor; And 4), the freedom to distribute copies of your modified versions to others.

The free software idea did not immediately become mainstream, and industry was especially suspicious of it. In 1998, the 'hacking' activists Bruce Perens and Eric Raymond agreed that a significant part of the problem resided in Stallman's term 'free software', which might understandably have an ominous ring to the ears of business people. Accordingly they, along with other prominent hackers, founded the 'open-source' software movement (Perens 1999). "Open-source" software incorporates essentially the same licensing practices as those pioneered by the free software movement. It differs from that movement primarily on philosophical grounds, preferring to emphasize the practical benefits of such licensing practices over issues regarding the moral rightness and importance of granting users the freedoms offered by both free and open-source software (von Hippel and von Krogh 2003).

From an innovation studies perspective, Lakhani and von Hippel (2003) and von Hippel (2005) suggested that open-source software development shows that users program to solve their own as well as shared technical problems and freely reveal their innovations without appropriating private returns from selling the software. Such "free" user-to-user assistance has turned open-source into a remarkable example of user-innovation (von Hippel 2009). It was also reported that the open-source trend has been so strong that previous, rather monolithic, organizations (e.g., SAP, Intel, Apple, Philips, Xerox, and IBM among others) decentralized research labs, open up proprietary technology, and increased their absorptive capacity for outside-in innovation processes within open-source ecosystems (Chesbrough et al. 2006; Enkel et al. 2009; Gassmann et al. 2010).

The open-source software phenomenon keeps evolving from the earliest purist views focusing on freedom (Stallman 1985) to newer perspectives considering open-source as an alternative and viable way of doing business (Ågerfalk and Fitzgerald 2008; Feller and Fitzgerald 2002; Fitzgerald 2006). Moreover, the phenomenon has expanded from open-source software to open-data (Gurstein 2011; Janssen et al. 2012), open-hardware (Maharaj et al. 2008; Söderberg 2013), open-knowledge (Awazu and Desouza 2004), open-access (Antelman 2004; Davis et al. 2008; Swan 2007), and open-medicine (Bradner 2011; Open Medicine Institute 2015), among other manifestations of increasing openness in the way of doing things. Even if the open-source phenomenon started to attract early scholarly attention in computer science and software engineering, the phenomenon is more recently capturing the largest interest from business and management scholars (Raasch et al. 2013). Therefore, as pointed out by Carillo and Bernard (2015); von Krogh and Spaeth (2007), information systems as a discipline is well positioned to be at the center of trans-disciplinary research addressing the phenomenon.

## III. Empirical background

### The cloud computing industry

Since our paper is about cooperation and competition, it is important to mention that the cloud computing business is dominated by a small number of players, including 1) Amazon, a pioneer in cloud computing services selling the Amazon EC2; 2) Google, selling services around its Compute Engine (Google Compute); and 3) Microsoft, heavily marketing cloud strategies based on its Azure cloud computing infrastructure (Microsoft Azure). Amazon, Google, and Microsoft do not provide cloud infrastructure products, merely computing services. In practice and if there were no alternatives, all cloud computation would run in hardware and software infrastructures controlled by very few players. Such control from the cloud computing service provider locks-in its customers (Armbrust et al. 2010; Briscoe and Marinos 2009; Chow et al. 2009). Surprisingly, the leading product alternatives to Amazon EC2, Google Compute and Microsoft Azure are not commercial but rather four open-source projects. They include: 1) OpenStack, our unit of analysis; 2) CloudStack, backed by Citrix and the Apache Software Foundation; 3) Eucalyptus, a system that is compatible with Amazon EC2 services and backed by many IT consulting firms; and 4) OpenNebula, more present in the European markets and backed by C12G, a Spanish company.

The three-dimensional Figure 1 provides a competitive overview of the key socio-technical ecosystems leading the cloud computing industry. As open-source technology can be freely used, studied, modified and distributed (Stallman 1985), we modeled the open-source ecosystems as wireframe spheres; and as proprietary technology relies on strong intellectual property protections, is less transparent and often developed behind corporate doors (Goldman and Gabriel 2005) we modeled proprietary ecosystems as opaque spheres. The size of the spheres took into consideration each ecosystem market-share.

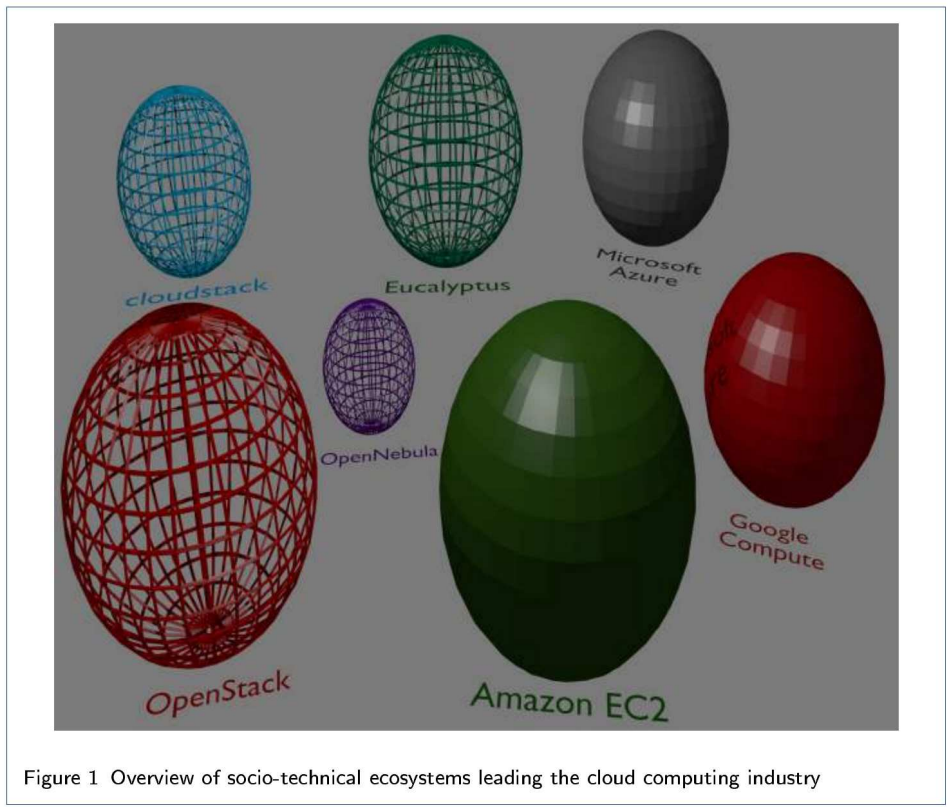

Figure 1 Overview of socio-technical ecosystems leading the cloud computing industry

## The OpenStack case

OpenStack is an open-source software cloud computing infrastructure capable of handling big data. It is primarily deployed as an "Infrastructure as a Service" (IaaS) solution. It started as a joint project of Rackspace, an established IT web hosting company, and NASA, the well-known U.S. governmental agency responsible for the civilian space program, aeronautics and aerospace research. Today, more than 200 firms and many non-affiliated individuals contributors pitch in to a set of different open-source projects governed by the OpenStack Foundation[2].

Both private companies (e.g., AT&T, AMD, Canonical, Cisco, Dell, EMC, Ericsson, HP, IBM, Intel, and NEC, among many others) and public entities (e.g., NASA, CERN, Johns Hopkins University, Instituto de Telecomunicações, Universidade Federal de Campina Grande, and Kungliga Tekniska Högskolan, among others) work together with independent, non-affiliated developers in a scenario of pooled R&D in an open-source way (i.e., emphasizing development transparency while giving up intellectual property rights). We decided to address the OpenStack case due to its perceived novelty, its high inter-networked nature (i.e., an "ecosystem" involving many firms and individual contributors), its heterogeneity (i.e., an ecosystem involving both startups and high-tech corporate giants), its market-size ($1.7bn, by 2016[3]), its complexity (i.e., involving different programming languages, different operating systems, different hardware configurations) and its size (17,020 community members, 100,000 code reviews, and 1,766,546 lines of code[4]).

Even if OpenStack emphasizes cooperation in the joint-development of a large open-source ecosystem, there are many firms directly competing with each other within the community. Among others, there is competition among providers of public cloud services based on OpenStack (e.g., HP, Canonical, and Rackspace), among providers of specialized hardware complementing OpenStack (e.g., HP, IBM, and Nebula), and among providers of complementary commercial software plug-ins complementing OpenStack (e.g., VMware, Citrix, and Cisco)[5].

## IV. Method

Our case relied on naturally occurring data which emerged *per se* on the Internet. Such data are not a consequence of researchers' own actions, but rather are created and maintained by the OpenStack community in their own pursuits of developing an open-source infrastructure for big data. In order to make sense of cooperative and competitive behaviors within OpenStack, we have combined qualitative analysis of archival data (QA), mining software repositories (MSR), and Social Network Analysis (SNA) on publicly-available and naturally-occurring data from the OpenStack Nova repository to reconstruct and visualize the evolution of cooperation in a sequence of networks. Table 2 presents the set of multidisciplinary methodological notes that guided our mixed methods research design.

We began our research in a qualitative way by screening publicly available data such as company announcements, financial reports and specialized press reports, which allowed us to review an immense amount of online information pertaining to the competitive cloud computing industry. While taking into consideration established methodological notes that legitimate the use of archival data when studying a case (Dubé and Paré 2003; Eisenhardt 1989; Flynn et al. 1990; Gibbert et al. 2008; Shanks 2002; Yin 2011), we gained valuable insights from the industrial context surrounding the OpenStack community. After attaining a better understanding of the industrial cooperative and competitive dynamics, we extracted and analyzed the social network of the OpenStack Nova project by leveraging SNA (Brandes et al. 2013; Scott 2012; Wasserman and Faust 1994).

---

[2]From a legal perspective, the OpenStack Foundation is a nonprofit, non-stock "foundation" within the meaning of Section 501(c)(6) of the Internal Revenue Code of 1986, "Delaware Corporate Law".
[3]http://451research.com/report-short?entityId=82593
[4]http://opensource.com/business/14/6/openstack-numbers
[5]For a relational map of competition among OpenStack firms (see Teixeira et al. 2015).

| Employed approach | Established within | Methodological notes |
|---|---|---|
| Case study[a] rooted on archival data | Multidisciplinary | Eisenhardt (1989)<br>Yin (2011)<br>Dubé and Paré (2003) |
| Mining software repositories | Software-Engineering | Kagdi et al. (2007)<br>Martinez-Romo et al. (2008)<br>Teixeira et al. (2015) |
| Social network analysis | Multidisciplinary | Freeman (2005)<br>Wasserman and Faust (1994)<br>Kane et al. (2014) |
| Network analysis of digital trace data | Information-Systems<br>Software-Engineering | Hahn et al. (2008)<br>Howison et al. (2012)<br>Trier (2008) |
| Network analysis with emphasis on the visualization of cooperative activities | Biomedicine<br>Bibliometrics<br>Innovation-Studies | Lundvall (1992)<br>Cambrosio et al. (2004)<br>Teixeira et al. (2015) |

[a] We see case study as an umbrella term covering a group of research methods which focus on a particular instance (see Adelman et al. 1976). Our case study depended on the use of – and ability to integrate in converging fashion (some would say "triangulate") – information from multiple sources of evidence. The evidence included announcements, financial reports, specialized press reports and actual information systems artifacts. To built around the consistency of complex data, we were forced to encapsulate different qualitative, quantitative and social network analysis methods (see Ibarra et al. 2005; Yin 1997). After all, "Using mixed methods within the confines of a single study can simultaneously broaden and strengthen the study. ... The stronger the "mix" of methods throughout these procedures the more that researchers can derive the benefits from using mixed approaches." (Yin 2006).

Table 2 Employed multidisciplinary methodological guidelines

As in Teixeira et al. (2015), we took advantage of naturally occurring digital trace data (i.e., the OpenStack Nova project repository and its *changelog*) and built cooperative social networks that were explored using a variety of tools: *Gephi* (v0.8.2) (Bastian et al. 2009), *Visone*(v2.7.3) (Brandes and Wagner 2004), and the *sna* (v2.3-2) and *statnet* (v2014.2.0) statistical modules (Butts 2010; Handcock et al. 2003) for *R* (v3.0.2) (R Core Team 2014). To better explore cooperation at the ecosystems level, we also modeled cooperative relationships in the tri-dimensional (3D) space using *Blender* (2.72). By mining digital traces of code cooperation, and by uncovering the social structure of the OpenStack Nova project, the computerized SNA also revealed key preliminary understandings of coopetition in the OpenStack ecosystem that were later re-investigated with complementary qualitative data. The combination of methods was not only fundamental for the retrieval of social structures, but also for explaining and them.

As in prior multi-disciplinary studies (Cambrosio et al. 2004; Glänzel and Schubert 2005; Lundvall 1992; Porter et al. 2005; Teixeira et al. 2015), our analysis emphasizes the visualization of the cooperation network, which evolves over time, to reveal the dynamics among the OpenStack software developers. We then attempted to understand the visualized networks with our acquired understanding from the competitive cloud computing industry in general and OpenStack in particular. The visualization, together with a deeper understanding of the phenomenon under investigation, corresponds to the notion of figuration (Elias 1978; Elias and Jephcott 1982; Smith 2001) as pointed out in several studies (Cambrosio et al. 2004; Gfaller 1993; Newton 2001).

Studies that take an SNA approach are long established in organizational research in general (Cross et al. 2002; Kane et al. 2014; Rowley et al. 2005; Tichy et al. 1979; Tsai 2002; Uzzi 1997), and in open-source research in particular (Crowston and Howison 2005; Geipel and Schweitzer 2012; Martinez-Romo et al. 2008; Zanetti et al. 2013). Even so, few studies have exploited the potential of social network visualizations for exploratory research, as recommended by Freeman (2005). By capturing the evolution of cooperative and competitive behaviors at the individual, firm and ecosystem levels, our multi-method approach aims to contribute to a broader range of methodologies capable of bridging the gap between causal processes at the macro-structural organization and those operating at the individual level (see Ibarra et al. 2005). We also argue, that our SNA approach integrating quantitative and qualitative evidence contributes to bridging the qualitative-quantitative divide in information systems research (see Ågerfalk 2013; Oinas-Kukkonen et al. 2010; Venkatesh et al. 2013).

## V. Results

We present our results in a narrative format complemented with "pictures" of the evolving social structure of OpenStack. Such narrative, concerned with how actors simultaneously cooperate and compete in the development of an open-source socio-technical ecosystems, aggregates theoretical issues that are later addressed in the discussion section. Such narrative, built on both quantitative and qualitative evidence, contributes to the understanding of a rather new and unexplored phenomenon and its embedded paradoxes (i.e., open-coopetition).

We start by quoting the words of Jim Curry in one of the first announcements of the OpenStack project. The founding leader of the OpenStack community starts by advocating the freedom of open-source software before stating the mission of the the OpenStack project. It is important to notice that Jim Curry emphasizes the roles of NASA and Rackspace as initial contributors to the project – that is, the project did not start from "ground zero".

> "What is OpenStack? Well, our mission statement says this: *To produce the ubiquitous Open Source Cloud Computing platform that will meet the needs of public and private clouds regardless of size, by being simple to implement and massively scalable.* That is a big ambition. The good news is that OpenStack is starting with code contributions from two organizations that know how to build and run massively scalable clouds – Rackspace and NASA. — Jim Curry, OpenStack Lead, 19 July 2010[6]

---

[6] http://www.openstack.org/blog/2010/07/introducing-openstack/

Visualizations in Figures 2 to 9 provide an understanding of how key players in the cloud computing industry cooperate in an open-source ecosystem. The size of a node is dependent on its degree-centrality – in other words, the larger the node, the more social connections the developer has. The value of degree-centrality depends on the number of adjacent nodes with which a node is connected. Therefore, the higher a developer's degree-centrality, the more likely he/she is to be cooperating with others.

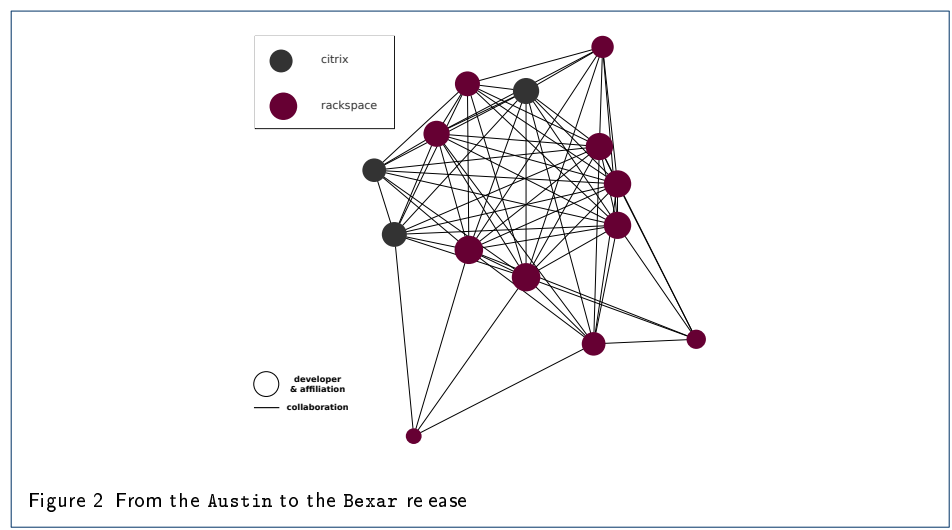

Figure 2 From the Austin to the Bexar release

Figure 2 captures the cooperation in the OpenStack Nova project from the `Austin` to the `Bexar` release, from October 21$^{st}$ 2010 to February 3$^{rd}$ 2011. From it, we can derive the cooperation between software developers affiliated with companies; for example, Citrix had three developers working on the project together with Rackspace.

> "OpenStack provides a solid foundation for promoting the emergence of cloud standards and interoperability." …. "As a longtime technology partner with Rackspace, Citrix will cooperate closely with the community to provide full support for the XenServer platform and our other cloud-enabling products." — Peter Levine, SVP and GM, Citrix, 19 July 2010 [7]

Citrix who had worked before with Rackspace, wanted to make sure that their XenServer platforms would be conveniently integrated with Rackspace's plans for OpenStack.

> "The project is exhibiting the key benefits that the industry derives from successful open-source cooperation: rapid development, faster testing, feedback and project turn around, broader industry adoption and learning through implementation and de-facto standardization whilst avoiding the prospect of commoditization.

---

[7] http://www.rackspace.com/blog/newsarticles/rackspace-open-sources-cloud-platform

It has been rewarding to work with the OpenStack crew, and to have experienced first hand the dedication to an open, code-rules, community-first approach taken by the project leaders." — Simon Crosby, CTO, Citrix 21 October 2010 [8]

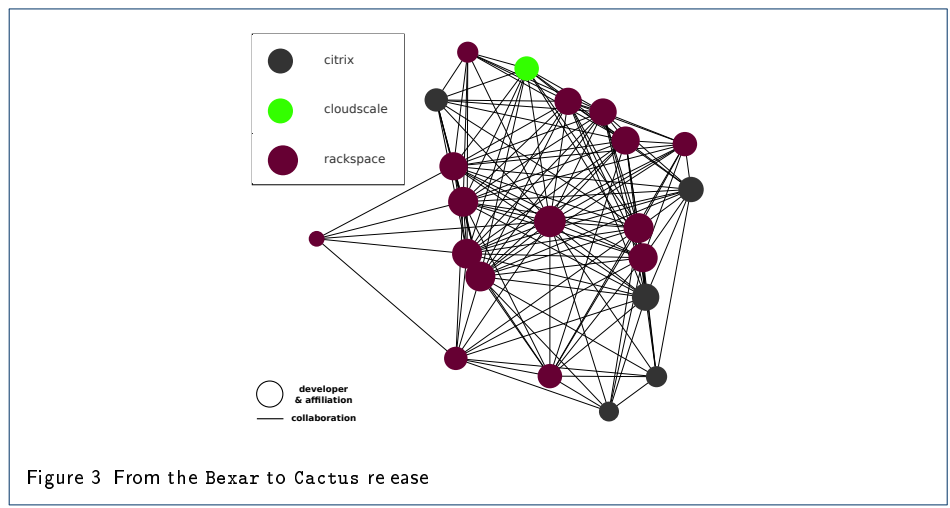

Figure 3 From the Bexar to Cactus re ease

Our second visualization with degree-centrality, in Figure 3, captures the cooperation from the `Bexar` to the `Cactus` release (from February 3$^{rd}$ 2011 to April 15$^{th}$ 2011). From this visualization we can observe a new node, a developer affiliated with Cloudscaling. Cloudscaling was founded in 2006 by the cloud architect and open-source software advocate Randy Bias and co-founder Adam Waters. It started as a professional services company selling custom cloud infrastructure for large service providers. They had 'Korea Telecom' as an early customer, for which the company in 2010 designed and deployed the first OpenStack-based storage cloud outside Rackspace.

"Earlier this week, one of our clients, a Tier 1 ISP, launched an object storage cloud based on OpenStack, an open-source compute and storage framework created by Rackspace and NASA. The new storage cloud is the first commercial OpenStack-based storage offering in the market after Rackspace itself, which is based on the same technology." – Joe Arnold, Director of engineering, Cloudscaling, 31 of January 2011[9]

Our visualization in Figure 4 captures cooperation from the `Cactus` to the `Diablo` release (from April 15$^{th}$ 2011 to September 22$^{nd}$ 2011). HP (a well-known IT multinational company), Mirantis (an OpenStack startup), and Red Hat (the company behind the Red Hat Enterprise Linux and sponsor of the Fedora Linux distributions) joined the coopetitive software development efforts.

---

[8] http://blogs.citrix.com/2010/10/21/if-youve-seen-one-redwood-youve-seen-them-all/
[9] http://cloudscaling.com/blog/cloud-computing/openstack-object-storage-moves-beyond-rackspace/

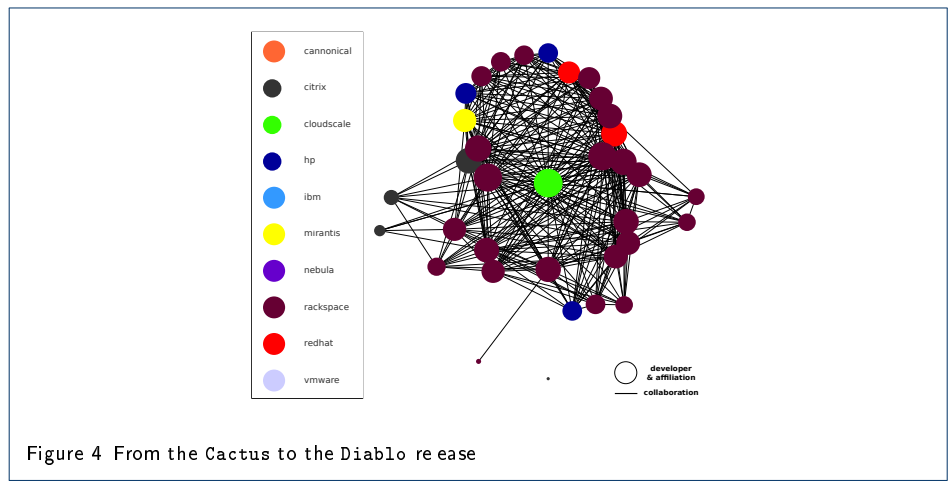

Figure 4 From the `Cactus` to the `Diablo` release

Mirantis was founded in January of 2011 by Boris Renski Jr. and Alex Freedland. Also born in Northern California, this startup marketed itself as a "pure-play" OpenStack company and started working early with Red Hat. Besides cooperating in the development of OpenStack, both firms partnered in implementation and integration services at common customers[10].

In the meantime, HP started marketing their cloud computing services based on OpenStack. HP markets itself as the leading corporation behind the project, emphasizing that OpenStack is the only cloud computing solution without a single-vendor lock-in but with an extensive ecosystem behind it[11].

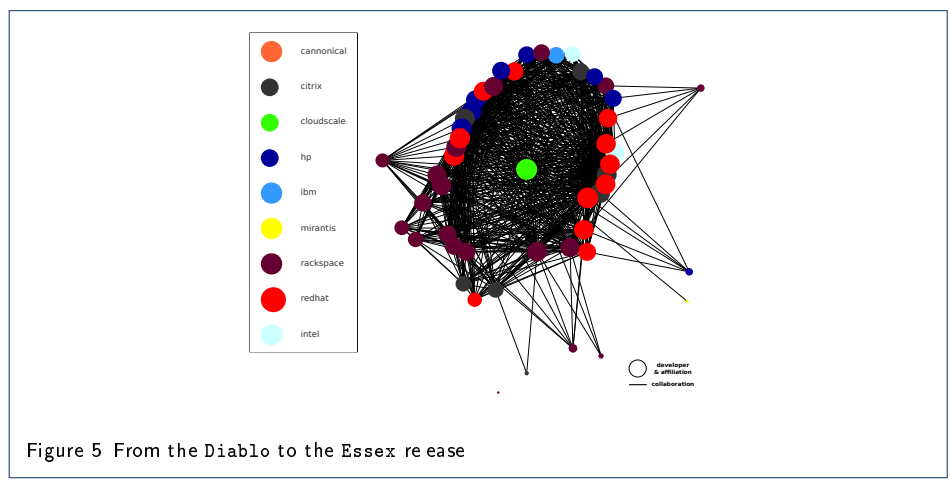

Figure 5 From the `Diablo` to the `Essex` release

Figure 5 depicts the cooperation from the `Diablo` to the `Essex` release (from September 22$^{nd}$ 2011 to April 5$^{th}$ 2012). Although the graph becomes more dense, we can visualize new nodes representing early contributions of Intel (investing on the compatibility of OpenStack with Intel microprocessors) and IBM. The later had a long history of working with open standards and open-source initiatives, such as in the Apache and Eclipse projects, and has been able to sell complementary solutions (i.e., hardware, software, and services) from open-source projects. It expected the same business model to work well with OpenStack[12].

---

[10]http://www.redhat.com/en/about/press-releases/red-hat-and-mirantis-partner-across-products-and-services
[11]http://www8.hp.com/h20621/video-gallery/us/en/events/enterprise/hp-discover-2011/2793799141001/

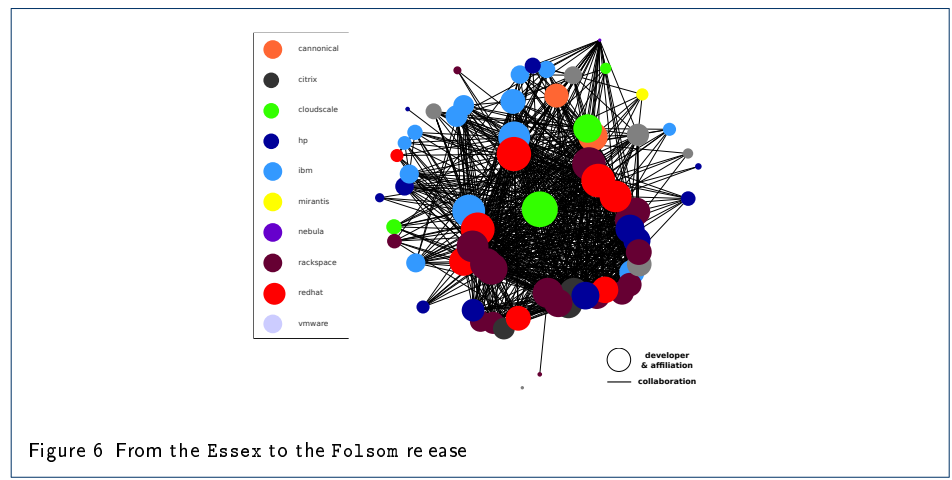

Figure 6 From the `Essex` to the `Folsom` release

Figure 6 shows the cooperation between the `Essex` and the `Folsom` release (from April 5$^{th}$ 2012 to September 27$^{th}$ 2012). We can observe that the network becomes more dense, as there are more developers working with each other. Even though some of their developers continued contributing to the project, Citrix had by then abandoned its OpenStack distribution in order to focus instead on the competing CloudStack cloud computing open-source ecosystem. Citrix decided to contribute to the competing CloudStack software ecosystem under the umbrella of the Apache Software Foundation, with a codebase resulting from the acquisition of Cloud.com of July 2011. This turn of strategy from Citrix was related with OpenStack's lack of integration with the Amazon's APIs (Application Programming Interfaces). Amazon is currently the leader of cloud services, and the migration costs to another cloud computing infrastructure are very high, specially if the APIs do not resemble each other.

> "Amazon has in many ways invented and created this market, and with what is projected to be $1 billion in ecosystem and customer revenue attached to Amazon cloud, we believe the winning cloud platform will have to have a high degree of interoperability with Amazon" – Sameer Dholakia, GM Cloud Platforms Group, Citrix, 3 of April 2012[13]

> "CloudStack has firmly aligned itself with the Amazon ecosystem. But OpenStack is an interesting case of an organization caught in the middle. Its service provider supporters are fundamentally interested in competing against

---

[12] http://thoughtsoncloud.com/2012/09/openstack-poised-to-lead-the-way-to-open-cloud-computing/
[13] http://www.networkcomputing.com/cloud-infrastructure/amazon-apis-are-fine--for-amazon/

AWS[14] ... they're afraid of a world in which AWS becomes the primary way that businesses buy infrastructure. It is to their advantage to have at least one additional successful widely-adopted cloud management platform in the market, and at least one service provider successfully competing strongly against AWS. Yet AWS has established itself as a de facto standard for cloud APIs and for the way that a service "should" be designed." – Lydia Leong, VP Distinguished Analyst, Gartner, 6 of April 2012[15]

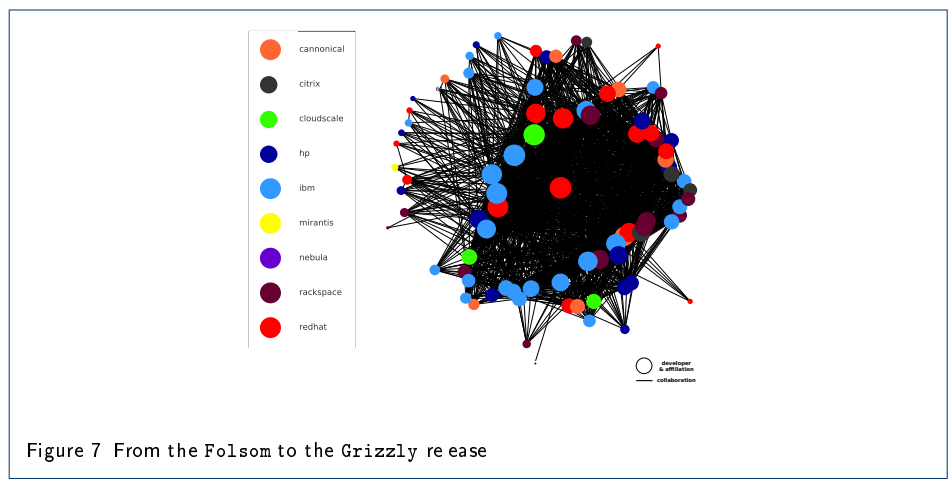

Figure 7 From the `Folsom` to the `Grizzly` release

Figure 7 shows the cooperation in the OpenStack Nova project from the `Folsom` to the `Grizzly` release (from September 27$^{th}$ 2012 to April 4$^{th}$ 2013). As expected, Citrix reduced its commitment to OpenStack, as we observe reduced cooperative activity from Citrix developers. Canonical continued investing increasingly in the development of OpenStack, interested in keeping its Linux Distribution *Ubuntu* as the leading Linux distribution for OpenStack clouds [16].

VMware, a Northern Californian firm with expertise in virtualization technologies, made substantial contributions (evidenced by the source-code commits) during this between-releases period[17]. The acquisition of the networking virtualization startup Nicira in July of 2012 reshaped the VMware cloud computing strategy. As a sign of commitment to OpenStack, VMware and Canonical issued a joint statement on their intentions to work together to improve the integration VMware technologies with Canonical's OpenStack distribution[18].

Figure 8 captures the cooperation in the project in a more recent phase, from the `Grizzly` to the `Havana` release (from April 4$^{th}$ 2013 to October 17$^{th}$ 2013). We can see that VMware took its commitment to OpenStack seriously, as six new developers engaged in developing with other OpenStack developers. Mirantis, in yellow on the right of Figure 8, invested heavily in cooperative activities with

---

[14]AWS stands for Amazon Web Services. See http://aws.amazon.com/ for more details.
[15]http://blogs.gartner.com/lydia_leong/2012/04/06/ecosystems-in-conflict-amazon-vs-vmware-and-openstak/
[16]http://www.markshuttleworth.com/archives/1373
[17]http://bitergia.com/openstack-releases-reports
[18]http://ir.vmware.com/releasedetail.cfm?ReleaseID=756729

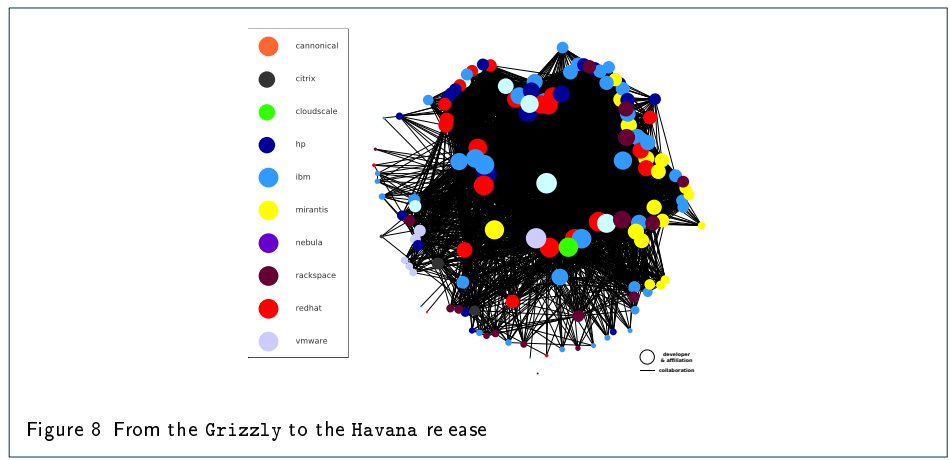

Figure 8 From the `Grizzly` to the `Havana` release

IBM, Rackspace, and Red Hat. Mirantis counted on financial support from Dell Ventures and Intel Capital (representing the interests of hardware manufacturers betting on OpenStack[19] and additional investment by Ericsson, Red Hat, and SAP Ventures[20] , turning it into one of the biggest code contributors to the OpenStack software ecosystem in just a few months[17].

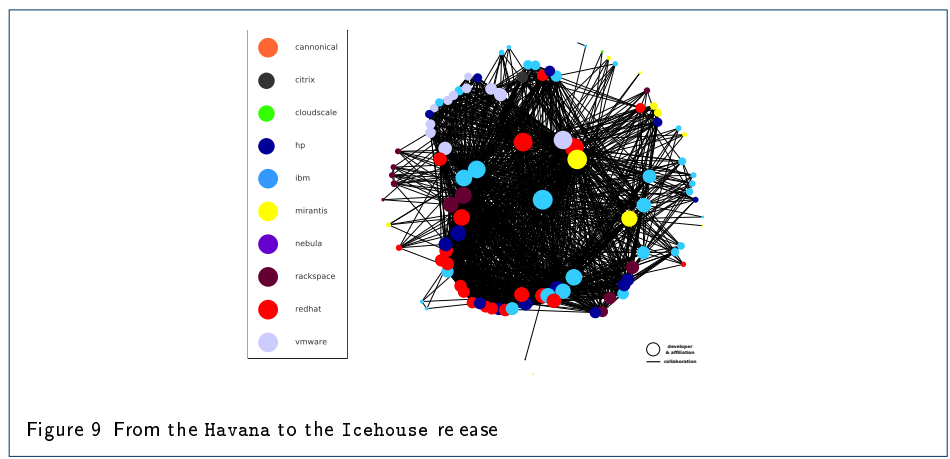

Figure 9 From the `Havana` to the `Icehouse` release

Figure 9 captures the cooperation in the latest period studied, from the `Havana` to the `Icehouse` release (from October 17[th] 2013 to April 17[th] 2014).

By this time the role of NASA on OpenStack had diminished. The first developments of OpenStack were in the service of science, supporting NASA's research activities. NASA's participation has been a selling point for advocates of OpenStack technologies. NASA lost much of its IT staff working on its Nebula cloud computing project. Software developers and IT architects headed to startups and high-tech giants within the OpenStack ecosystem. Moreover, a cost-driven IT reform led to disinvestment in OpenStack by NASA. Today, scientists at NASA depend on the Amazon EC2 and Microsoft Azure cloud computing infrastructures[21]. Meanwhile, on the other side of the Atlantic, the European Organization for Nuclear Research (CERN) decided on an OpenStack based strategy in 2012. In January of 2014, OpenStack was already running collision reconstructions at the Large Hadron Collider (LHC)[22].

---

[19]http://www.mirantis.com/company/press-center/company-news/mirantis-receives-10-million
[20]http://www.mirantis.com/company/press-center/company-news/mirantis-with-investment/

By pure serendipity, we also explored the cooperation among competing ecosystems (i.e., OpenStack vs. CloudStack). As pointed out before, Citrix, in a surprise move, and citing the lack of OpenStack interoperability with Amazon cloud systems, acquired Cloud.com in July of 2011 and donated the overall code-base to the Apache Foundation (a reputed non-profit corporation supporting open-source software). "CloudStack's application programming interface (API) provides compatibility with Amazon Web Services' Elastic Compute Cloud (EC2), the world's most popular public cloud" ... "(Citrix) hopes that EC2 customers will use their CloudStack for their private clouds while using the EC2 APIs to integrate with Amazon's public cloud"[23]. This move raised conflict in the OpenStack community. Citrix claimed to make peace with OpenStack on 21 April 2015 by announcing that it had become a Corporate Sponsor of the OpenStack Foundation[24].

From a research point of view, it was unclear during this period of conflict, whether Citrix developers were contributing to OpenStack, CloudStack or both. By mining both OpenStack and CloudStack repositories with SNA, we found out that 10 developers contributed both to OpenStack and CloudStack. Six of these developers were affiliated with Citrix. While Citrix's contributions were recurrent, the contributions of the other four were sporadic.

In the following Figure 10, we capture the role of Citrix in cooperation among competing ecosystems, from the first open-source release of CloudStack to Citrix's official return to OpenStack (from November 6th 2012 to April 21th 2015). We represent the ecosystems as wireframe spheres, OpenStack on the left and CloudStack on the right. Inside each ecosystem, we modeled the cooperative networks of developers that contribute to both ecosystems. As open-source ecosystems are not black boxes, we were able to identify such developers. Citrix, as a firm, is represented as a cuboid. The ultra-thin cylinder connecting firms with developers maps an "affiliation" relationship. Such visualization highlight that Citrix's developers were contributing to both OpenStack and CloudStack – in the open-source arena, developers and firms can take part in competing ecosystems.

## VI. Discussion

This research was conducted with the purpose of exploring cooperation among competitors in the open-source arena. Our case revealed certain peculiarities that call for the adaptation and expansion of the selected theory. Among other phenomenological peculiarities we highlight: (1) the inclusiveness of the OpenStack ecosystem to third-party contributors – this contrasts with literature derived from joint ventures and R&D consortia where access is granted only to a few selected members (cf. Büchel 2003; Sydow et al. 2012); (2) the openness and transparency of the OpenStack ecosystem – this contrasts with the literature on inter-organizational relations that emphasize gate-keeping and intellectual property rights (cf. Granstrand and Holgersson 2013; Pagani 2013); and (3), that actors within our narrative cooperate and compete simultaneously at the inter-individual, inter-firm and inter-ecosystem level as demonstrated by Citrix dual-role in both OpenStack and CloudStack – coopetition among ecosystems remains largely unexplored as pointed out in a recent review by Bengtsson and Kock (2014).

---

[21]http://blogs.nasa.gov/NASA-CIO-Blog/2012/06/09/post_1339205656611/
[22]http://home.web.cern.ch/about/updates/2014/01/importance-openstack-cern
[23]http://www.zdnet.com/article/openstack-vs-cloudstack-the-beginning-of-the-open-source-cloud-wars/
[24]http://www.citrix.com/news/citrix-in-the-news/apr-2015/citrix-makes-peace-with-openstack

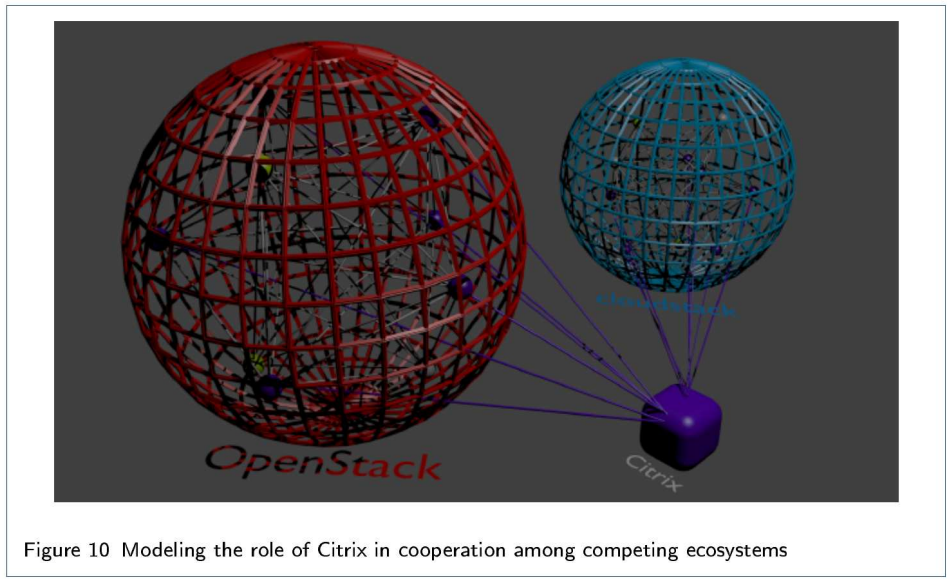

Figure 10 Modeling the role of Citrix in cooperation among competing ecosystems

From the perspective of a network of firms embedded in a socio-technical ecosystem, our research adds to the paradox literature by responding specifically to Lewis (2000) and to Poole and Van de Ven (1989), who encouraged researchers to construct concepts and theories dealing with paradoxical organizational complexities (Chen 2008). It pioneers by integrating paradigms on competition, cooperation, and open-source software. In this regard, we address Lewis and Grime's (1999) call for research converging disparate paradigms. Aligned with theory-building strategies proposed by Poole and Van de Ven (1989) to take advantage of paradoxes of social theory, we adopted multiple levels of analysis (such as individual, firm, dyad, and ecosystem) and theorized on the recent conception of "open-coopetition" proposed by Teixeira and Lin (2014) for a better understanding of how competing firms cooperate in an open-source way.

This research also contributes to prior research efforts in theoretically integrating the concepts of cooperation and competition (Ahuja 2000; Bengtsson and Kock 2000; Chen 2008; Chen and Miller 2015; Dyer and Singh 1998; Gnyawali et al. 2010; Gnyawali and Park 2011). More particularly, our case confirmed much of the established literature addressing the competitive-cooperative mode of interorganizational interactions where firms "cooperate with competitors to win" in a self-serving game to ultimately gain the upper hand (Brandenburger and Nalebuff 2011; Chen and Miller 2015; Hamel et al. 1989).

As theoretically expected, competing firms cooperated within the OpenStack socio-technical ecosystem seeking complementary external resources (Bengtsson and Kock 2000; Tsai 2002), to mitigate risk (Gnyawali and Park 2011) and to learn from others (Dussauge et al. 2000). Through cooperation, two companies can gain access to each other's unique resources or share the cost of developing new unique resources (Bengtsson and Kock 2000). Within an open-source scenario, it is an open (i.e., open for contributions from everyone) and networked community (i.e., relating firms and independent developers) that endows such external resources. Even if the established competition theory did not originate from the open-source context and rarely takes information systems into account in its conceptualizations, its theoretical lenses proved to be very useful to make sense of why a network of organizations would joint develop a cloud infrastructure for big data in an open-source way.

However, our research also contrasts with the established competitive-cooperative literature. First, by the inclusive and transparent nature of open-source projects – most research addresses alliances that take the form of joint ventures, consortia, or other arrangements where access is granted only to a few selected partners (Bengtsson and Kock 2014; Gnyawali and Madhavan 2001). As in other open-source projects, in OpenStack, everyone is welcome to contribute, and everyone is allowed to copy, sell, and distribute outcomes from the project. In the OpenStack case, cooperation is wider, more heterogeneous, and more networked than in much of the competitive-cooperative literature. The cooperation in OpenStack included research institutes (e.g., CERN and NASA among many others), universities (e.g., Johns Hopkins University, and Kungliga Tekniska Högskolan among many others), clients (e.g., Korea Telecom, and AT&T among many others deploying OpenStack), hardware suppliers (e.g., Intel and Cisco), competitors (e.g., IBM and HP), and many independent third-party developers which are often hobbyists without clear affiliation. Future research is needed to assess how the nature of open-source software influences coopetitive relationships.

We conducted the analysis of relational interactions at the inter-individual, inter-firm and inter-ecosystem level. As research addressing coopetition among ecosystems remains scarce (Bengtsson and Kock 2014), we centered our discussion on results that explore OpenStack in relation to its competing ecosystems (i.e., CloudStack and Amazon). We highlight the role of Citrix, a key actor in the ecosystem which commitment to OpenStack evolved over time. As reported earlier, by 2010, Citrix was the partner of Rackspace on the early days of OpenStack at NASA (see Figure 2). On the follow-up of Citrix's acquisition of Cloud.com on July 2011, Citrix announced to 'abandon' OpenStack in September 2012 citing the lack of integration with the Amazon's APIs. Citrix opted to focus instead on CloudStack which codebase was open-sourced with its acquisition of Cloud.com (see Figures 5 to 8 for the progressively decreasing presence of developers affiliated with Citrix). Citrix claimed later to make peace with OpenStack on 21 April 2015 by official becoming a Corporate Sponsor of the OpenStack. By mining both OpenStack and CloudStack repositories, we found out that six developers affiliated with Citrix were actually recurrently contributing to both competing ecosystems during the conflict period (see Figure 10 ).

From the perspective of a dyad of socio-technical ecosystems, we add to literature on portfolio of alliances (Anand and Khanna 2000; Duysters et al. 2012; Hoffmann 2007; Lavie 2007; Lavie and Miller 2008; Lavie and Singh 2012; Wassmer 2010). As highlighted in Figure 10, we remarked that developers affiliated with Citrix contributed concurrently to two competing open-source ecosystems (i.e., OpenStack and CloudStack). In fact, Citrix recurrently engaged in cooperation with many other IT giants (e.g., with Amazon in setting cloud computing interoperability standards[25], with Microsoft in 'PC' virtualization technologies[26] and with Google by joint-developing software powering the 'Chromebook laptop family'[27] among others). Such netting cooperative behavior of Citrix was not exceptional, HP also contributed to both OpenStack and Eucalyptus (another open-source cloud computing ecosystem). The observed mesh of alliances adds relevance to research on how to manage a portfolio of alliances (Hoffmann 2005,0). Anand and Khanna (2000) argued that firms should build an alliance management capability while Hoffmann (2005) reported on the institutionalisation of multi-alliance management.

Research on multi-alliance management addressed the effects of alliance portfolio diversity. However, results on a diversity-performance relationship are rather inconclusive with mixed results (Duysters et al. 2012). It is generally accepted that redundant alliances can increase the reliability of the information and resources that a focal company has access to and also reduce the dependence on a single partner (Hoffmann 2005). However, the advantage of a portfolio of alliances is not so much a matter of the portfolio's size (i.e. number of alliances), but a matter of characteristics of the firms that a focal organization is connected to (Stuart 2000). Also, that the performance of a focal firm improves with the intensity of competition among partners in its alliance portfolio (Lavie 2007). The evolution of alliance portfolios is contingent on external constraints and opportunities (changes in exogenous uncertainty), internal resources available, and strategic choices that interact and drive the pace, pattern, and direction of the evolution of a firm's alliance portfolio (Hoffmann 2007). Inertial pressures tend to fixate the current configuration of the alliance portfolio, whereas external stimuli in the form of technological changes and shifts in strategy set new courses for alliances (Lavie and Singh 2012).

---

[25]http://www.citrix.com/global-partners/amazon-web-services/overview.html
[26]http://www.citrix.com/global-partners/microsoft/overview.html
[27]http://www.citrix.com/news/announcements/aug-2014/citrix-collaboration-with-google

As information systems become increasingly networked and interconnected (Ciborra et al. 2000; Henfridsson and Bygstad 2013) and research in multi-alliance management remains scarce and inconclusive (Duysters et al. 2012; Lavie and Singh 2012), there is an opportunity for information systems research to bridge the literature in multi-alliance management and the literature in open-source software. Given the trans-disciplinary nature of information systems (Carillo and Bernard 2015; Galliers 2003) the IS field would be a perfect candidate for engaging in a trans-disciplinary dialog and play a central role in producing a cumulative body of high-quality research grasping both multi-alliance management (Rai and Tang 2010) and open-source software (von Krogh and Spaeth 2007). As become salient in our case by the pivotal role of Citrix across ecosystems, and as evidenced by the sampled cases of open-coopetition (see Table 1), firms are often involved in multiple and often competing open-source ecosystems. However, it remains largely unknown how organizations manage their involvement across multiple and often competing open-source ecosystems – a promising avenue for future research.

A wide range of research focused on competition between ecosystems Apple Safari vs. Microsoft Internet Explorer, Nokia N-Gage vs. Nintendo Gameboy, Microsoft Music Player vs. Apple iTunes, Google Android vs. Apple iPhone and Google Talk vs. Skype as captured in a recent study by Eisenmann et al. (2011). Researchers tend to stress on ecosystems as competing entities while ignoring cooperation among them. By exploring both competition and cooperation among ecosystems, we propose that development transparency and the weak intellectual property rights, two well-known characteristics of open-source ecosystems, allow an easier transfer of information and resources from one alliance to another (see Hoffmann 2005). The scholars Parise and Casher (2003) pointed out that "sharing of information and knowledge across these alliances is rare", but Hoffmann (2005) pointed out synergies from transferring information and resources from one alliance to another. Future research is needed to access the impacts of development transparency and weak intellectual property rights on spillovers, after all, open-source initiatives are "free spillovers" (West and Gallagher 2006) or "spillovers without compensation" (Gassmann et al. 2010).

By exploring the OpenStack ecosystem, following especially the evolution of Citrix's competitive-cooperative actions, we propose the following theoretical proposition pinpointing a peculiarity of cooperating with competitors in an open-source way:

**Theoretical Proposition 1** – *Within an R&D context where a focal firm is engaged in multiple alliances, development transparency and weak intellectual property rights, allow an easier transfer of information and resources between alliances.*

At first sight, and while providing insight into why firms cooperate with competitors in an open-source way, the advanced proposition makes a virtue of open-coopetition (i.e., because it enables a focal firm to transfer information and re-sources more easily between multiple alliances). However, living with paradox implies that we shift expectations for rationality and linearity to accept paradoxes as persistent and unsolvable puzzles (Smith and Lewis 2011). Consequently, we caution against seeing the "ease of transfer" as a pure virtue. If in one hand, the focal firm can benefit from an easier transfer of information and resources between alliances, on the other hand, its competitors can also benefit from transparency and weak intellectual rights (either entering in the same alliances or not). Even within the context of multiple alliances, open-coopetition remains paradoxical and difficult to explain. From one side, organizations can simultaneously benefit from both coopetition (see Bengtsson and Kock 2000; Chen and Miller 2015; Gulati et al. 2012) and open-source innovation (see Ågerfalk and Fitzgerald 2008; Fitzgerald 2006; von Hippel and von Krogh 2003). But from the other side, there is an increased risk of opportunistic behavior and unintended spillover effects (see Nooteboom et al. 1997; Park and Russo 1996; Trott and Hartmann 2009).

Given the specific empirical background, where OpenStack challenges the dominance of the cloud computing market by three large players (i.e., Amazon, Google, and Microsoft), we add OpenStack as yet another paradoxical case of firms' investment in open-source software to disrupt the leading positions of proprietary software players (see West 2003; West and Gallagher 2006). However, as pointed out in a recent review by Teixeira and Baiyere (2014), "tables turned", open-source ecosystems should no longer be seen as the alternative to proprietary ecosystems controlled by leading, and often monopolistic, players (e.g., Linux vs. Windows, Mozilla vs. Internet Explorer, and R vs. SPSS among others). As outlined in Table 1, several open-source ecosystems are now leading the market. For example, Google Android and the Open Handset Alliance are leading the mobile platforms market while traditional proprietary software players, such as Microsoft and Blackberry, are currently struggling with residual sales (Teixeira and Baiyere 2014).

We remain far from a comprehensive understanding of "why, how, and when competitors cooperate in open-source ecosystems". However, our research efforts suggest that coopetition in an open-source way have its peculiarities - future research is needed to assess the impact of the open-source movement to theory addressing both cooperation and competition in inter-organizational settings.

## VII. Conclusion

Managers deal with tensions in which opposing forces push or pull the organization in several ways at the same time (Lewis 2000; Nutt and Backoff 1993). In industries characterized by high R&D costs and uncertainty (e.g., information technology sector), managers might have to deal with the *competition versus cooperation* and *open-source versus proprietary* technology development paradoxes simultaneously. Both by visualizing the evolution of the social structure of OpenStack and by scrutinizing complementary qualitative data, we propose that the development transparency and weak intellectual property rights (i.e., characteristics of open-source ecosystems) allow a focal firm to transfer information and resources more easily between its multiple alliances.

# Acknowledgments

Besides the three anonymous reviewers, we would like to thanks participants from the 1st European Conference on Social Networks (INSNA EUSN 2014), the 10th International Symposium on Open Collaboration (ACM OpenSym 2014), the 2nd IEEE International Conference Big Data (IEEE Big Data2014) and the 38th Information Systems Research Conference in Scandinavia (IRIS38) for early feedback on our research efforts[28]. Special thanks to Gregorio Robles and Jesus Gonzalez-Barahona that brought much expertise from the Mining Software Repositories (MSR) field. Acknowledgments also for Lin Tingting, Devi Gnyawali, Rudy Hirschheim, Natalia Levina, Sonali Shah and Annabelle Gawer for valuable ideas, comments, and conversations. We thank also Arto Lanamäki for help revising the manuscript. This research was partially funded by the Fundação para a Ciência e a Tecnologia (grant SFRHBD615612009), Liikesivistysrahasto (grants 5-3076 and 8-4499), Marcus Wallenberg Säätiö (grant "open-coopetition R&D management strategy"), Academy of Finland (decision no. 295743) and the Turku Centre for Computer Science (travel grant). We thank also the University of Turku, the EIT digital and the University of Oulu for providing much of the needed research infrastructure. Finally, we would like to thank the R, Gource, Gephi, Tulip, Blender and OpenStack open-source communities for developing cool and research-friendly software.

**Author details**

[1]Turku School of Economics, University of Turku, Turun yliopisto, 20014 Turku, Finland. [2]OASIS Research Group, University of Oulu, Oulun yliopisto, 90014 Oulu, Finland.

---

[28]See Teixeira (2014a/2014b/2014c) as well as Teixeira and Mian (2015) for earlier related research.